\documentclass{article}
\usepackage{spconf}
\usepackage{{graphicx}}
\usepackage{amsmath}
\usepackage{amssymb}
\usepackage[linesnumbered,ruled]{algorithm2e}

\ninept

\title{Study of Robust Adaptive Power allocation for the Downlink of Multiple-Antenna Systems}

\name{Andre R. Flores and Rodrigo C. de Lamare}
\address{Centre for Telecommunications Studies (CETUC) \\
Pontifical Catholic University of Rio de Janeiro\\
Emails: delamare@cetuc.puc-rio.br
 and andre.flores@cetuc.puc-rio.br}

\begin{document}

\maketitle

\begin{abstract}
 Multiple-input multiple-output (MIMO) systems greatly increase the overall throughput of wireless systems since they are capable of transmitting multiple streams employing the same time-frequency resources. However, this gain requires an appropriate precoder design and a power allocation technique. In general, precoders and power allocation schemes are designed assuming perfect channel estate information (CSI). Nonetheless, this is an optimistic assumption since real systems only possess partial or imperfect CSI at the transmitter (CSIT). The imperfect CSIT originates residual inter-user interference, which is detrimental for wireless systems. In this paper, two adaptive power allocation algorithms are proposed, which are more robust against CSIT imperfections than conventional techniques. Both techniques employ the mean square error as the objective function. Simulation results show that the proposed techniques obtain a higher performance in terms of sum-rate than conventional approaches. 
\end{abstract}



%

\section{Introduction}

Modern wireless communications systems rely on architectures where both the transmitter and the receiver are equipped with multiple antennas, also known as multiple-input multiple-output (MIMO) systems \cite{mmimo,wence,Shafi2017}. The main advantage of MIMO systems is that they increase dramatically the overall throughput without the need for additional bandwidth \cite{Li2014}. This gain comes from the simultaneous transmission of multiple data streams which share the same time-frequency resources.

An efficient transmission over the downlink (DL) of a MIMO system depends on the appropriate design and implementation of precoding and power allocation schemes \cite{Spencer1,Stankovic,Sung,Zu_CL,Zu,wlbd,rsbd,rmmse,wlbf,locsme,okspme,lrcc,mbthp,rmbthp,rsthp,bbprec,baplnc,jpba,zfsec,rprec&pa,cgabd}. Precoding allows the decoding of the information at the receiver by exploiting the multipath propagation and suppressing the multiuser interference (MUI) \cite{deLamare2003,itic,spa,cai2009,jiols,jiomimo,mfdf,mbdf,did,rrmser,jidf,bfidd,1bitidd,dynovs,detection_review,aaidd,listmtc,dynmtc,mwc,dynovs,dopeg,memd,vfap}. In addition, the achievable data rate associated with the users is affected by the power allocation scheme adopted, which must allocate the power levels according to the channel conditions. 

The power allocation problem becomes non-convex in a multiuser MIMO (MU-MIMO) scenario due to the MUI, which requires an exhaustive search over the entire space of possible values to find the optimal. Indeed, the power allocation problem is an NP-hard problem \cite{Luo2008,Liu2011} and finding the optimal solution is computational demanding. In \cite{Utschick2012} the optimal power allocation is found through monotonic optimization at the expense of an exponential growing in computational complexity. In \cite{Schubert2004}, a scenario where the users are equipped with single-antenna terminals is analyzed. In \cite{Codreanu2007}, the previous work is extended to a MU-MIMO scenario where devices with multiple antennas are considered. Both approaches require the formulation and solution of geometric programming (GP) which is computational demanding. The connection between SINR and weighted sum rate (WSR) has been explored in \cite{Tan2011}. In \cite{Wang2015}, the receiver and the power allocation parameters of a multihop wireless sensor network are found through alternating optimization. Due to the complexity of the previous approaches, in \cite{Shi2011,Bjornson2011} local optimal solutions have been studied, reducing the complexity. Practical power allocation algorithms based on water-filling are presented in \cite{Palomar2005}, whereas algorithms based on the weighted minimum mean square error (WMMSE) minimization are reported in \cite{Christensen2008} and fractional programming (FP) \cite{Shen2018} are presented in \cite{Chakraborty2021}. In \cite{Palhares2020}, power allocation techniques based on convex optimization and adaptive processing were proposed for cell-free MIMO systems to maximize the minimum rate achieved among users.

In general, precoding and power allocation are performed assuming perfect knowledge of the channel state information at the transmitter (CSIT). Under this assumption the optimal power for diverse performance metrics (such as BER, sum-rate and fairness) is usually known \cite{Castaneda2017} and closed form expressions for specific setups are available, such as for zero-forcing precoders \cite{Bartolome2006}. However, time division duplex (TDD) systems employ training pilots to acquire CSIT whereas frequency division duplex (FDD) systems depend on feedback links. Both methods introduce error in the estimation procedure, which leads to an imperfect CSIT estimate \cite{Vu2007}. Thus, real world systems do not meet the perfect CSIT assumption. The imperfect CSIT originates residual MUI which is detrimental to the system performance. Robust precoding techniques based on the worst-case performance optimization have been proposed to deal with the uncertainties \cite{Vorobyov2003,Yu2010}. Later, the precoder design incorporates subspace projection techniques to increase the robustness against CSI imperfections \cite{Ruan2019}. Since CSIT imperfections degrade heavily the overall system performance, the design of robust techniques is of great importance. 

In this paper, two adaptive power allocation techniques are developed, namely the mean square error adaptive power allocation (M-APA) and the robust M-APA (RM-APA). These techniques minimize the mean square error (MSE) between the information at the transmitter and the received signal. The main difference between these techniques is that the RM-APA employs statistical information of the CSIT error. The performance achieved is compared with the uniform power allocation (UPA) and the optimal power allocation. Simulation results show that the APA algorithms attain a better performance in terms of sum-rate than UPA and comparable to optimal power allocation. 

The rest of this paper is organized as follows. In Section \ref{section system model}, the system model is presented. The M-APA and RM-APA algorithms are derived and detailed in Section \ref{section adaptive power allocation}. Section \ref{section Simulations} shows the simulation results, where the sum-rate of the proposed and conventional techniques are depicted. Finally, Section \ref{section conclusion} draws the conclusions of this work.

\section{System model}\label{section system model}

Let us consider the DL of a MU-MIMO system where the BS, which is equipped with $N_t$ antennas, transmits data to $K$ users. The $k$th user is equipped with $N_k$ antennas. Therefore, the total number of receive antennas is given by $N_r=\sum_{k=1}^{K}N_k$. The messages are encoded and modulated into a vector of symbols $\mathbf{s}^{\text{T}} \in \mathbb{C}^{N_r}$. A power allocation matrix $\mathbf{A} \in \mathbb{R}^{N_r\times N_r}$ contains the weights that allocates the power to the symbols. Once the power is allocated, a precoder $\mathbf{P} \in \mathbb{C}^{N_t\times N_r}$ maps the symbols to the transmit antennas \cite{Joham2005}. Then, the transmitted vector $\mathbf{x}\in \mathbb{C}^{N_t}$ can be expressed as follows: 
\begin{align}
\mathbf{x}=&\mathbf{P}\mathbf{A}\mathbf{s}=\mathbf{P}\text{diag}\left(\mathbf{a}\right)\mathbf{s}\nonumber\\
=&\sum_{m=1}^{N_r}a_m s_m \mathbf{p}_m. \label{Transmit Signal}
\end{align}
The system has a transmit power constraint given by $\mathbb{E}\left[\lvert\mathbf{x}\rvert^2\right]\leq E_{tr}$, where  $E_{tr}$ denotes the total available power.

Once the information is ready for transmission it is sent to the receivers through a channel $\mathbf{H}=\mathbf{\hat{H}}+\mathbf{\tilde{H}}\in \mathbb{C}^{N_r\times N_t}$. The matrix $\mathbf{\hat{H}}$ represents the channel estimate and the matrix $\mathbf{\tilde{H}}$ models the CSIT imperfection by adding the error of the estimation procedure. Each coefficient $h_{ij}$ of the matrix $\mathbf{H}$ represents the link between the $i$th receive antenna and the $j$th transmit antenna. The channel matrix can be expressed by $\mathbf{H}=\left[\mathbf{H}_1,\mathbf{H}_2,\cdots,\mathbf{H}_K\right]$, where $\mathbf{H}_k$ denotes the channel connecting the BS to the $k$th user.

The received signal obtained following the model established is 
\begin{equation}
\mathbf{y}=\mathbf{H}\mathbf{x}+\mathbf{n},\label{General Receive vector}\\
\end{equation}
where $\mathbf{n}\in\mathbb{C}^{N_r\times 1}$ is the additive noise modelled as a circularly symmetric complex Gaussian random vector, i.e. $\mathbf{n}\sim \mathcal{CN}\left(\mathbf{0},\mathbf{R_{nn}}\right)$.

\section{Adaptive Power Allocation}\label{section adaptive power allocation}

Let us consider the MU-MIMO model described in the previous section. Assuming knowledge of the precoder which remains fixed during the transmission of a packet, the problem is to find suitable values for the coefficients $a_i ~~\forall ~i=1,2,\cdots,N_r$ to enhance the overall performance of the system. For this purpose let us consider the minimum mean square error (MMSE) between the transmitted signal and the estimated signal at the receiver as the objective function given by
\begin{equation}
\begin{gathered}
\min_{\mathbf{a}} \mathbb{E}\left[\varepsilon\right]\\
\text{s.t.}~~\text{tr}\left(\mathbf{P}\text{diag}\left(\mathbf{a}\odot \mathbf{a}\right)\mathbf{P}^{H}\right)=\frac{E_{tr}}{\sigma^2_s},
\end{gathered}\label{obejctive function 1}
\end{equation}

where the error is defined as $\varepsilon=\lVert\mathbf{s}-\mathbf{y} \rVert^2$ and the transmit power constraint is $\frac{E_{tr}}{\sigma^2_s}$. Evaluating the error, we get
\begin{align}
\varepsilon=&\lVert \mathbf{s}-\mathbf{H}\mathbf{P}\text{diag}\left(\mathbf{a}\right)\mathbf{s}-\mathbf{n}\rVert^2\nonumber\\
=&\mathbf{s}^H\text{diag}\left(\mathbf{a}\right)\mathbf{P}^H\mathbf{H}^H\mathbf{H}\mathbf{P}\text{diag}\left(\mathbf{a}\right)\mathbf{s}\nonumber-\mathbf{s}^H\mathbf{H}\mathbf{P}\text{diag}\left(\mathbf{a}\right)\mathbf{s}\\
&-\mathbf{s}^H\text{diag}\left(\mathbf{a}\right)\mathbf{P}^H\mathbf{H}^H\mathbf{s}_p+\mathbf{s}^H\text{diag}\left(\mathbf{a}\right)\mathbf{P}^H\mathbf{H}^H\mathbf{n}\nonumber\\
&+\mathbf{s}^H\mathbf{s}+\mathbf{n}^H\mathbf{H}\mathbf{P}\text{diag}\left(\mathbf{a}\right)\mathbf{s}-\mathbf{n}^H\mathbf{s}-\mathbf{s}^H\mathbf{n}+\mathbf{n}^H\mathbf{n}.\label{Squared Error}
\end{align}

\begin{figure}[t!]
\begin{center}
\includegraphics[scale=0.4]{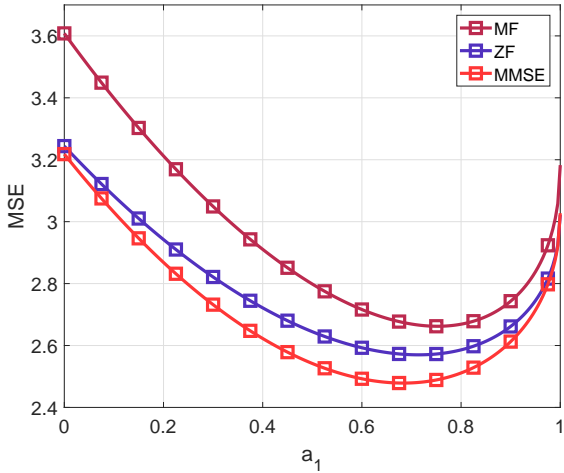}
\vspace{-1.5em}
\caption{Objective function: MSE with two streams}
\label{C5 T1}
\end{center}
\end{figure}

Remark that equation \eqref{Squared Error} is a scalar. Thus we can apply the trace operator over the right-hand side of the equation while preserving the equality. By applying the property $\text{tr}\left(\mathbf{C} + \mathbf{D}\right)=\text{tr}\left(\mathbf{C}\right)+\text{tr}\left(\mathbf{D}\right)$, where $\mathbf{C}$ and $\mathbf{D}$ are two general matrices with the same dimension, we obtain
\begin{align}
\varepsilon=&\text{tr}\left(\mathbf{s}^H\mathbf{s}\right)+\text{tr}\left(\mathbf{s}^H\text{diag}\left(\mathbf{a}\right)\mathbf{P}^H\mathbf{H}^H\mathbf{H}\mathbf{P}\text{diag}\left(\mathbf{a}\right)\mathbf{s}\right)\nonumber\\
&-\text{tr}\left(\mathbf{s}^H\mathbf{H}\mathbf{P}\text{diag}\left(\mathbf{a}\right)\mathbf{s}\right)-\text{tr}\left(\mathbf{s}^H\text{diag}\left(\mathbf{a}\right)\mathbf{P}^H\mathbf{H}^H\mathbf{s}\right)\nonumber\\
&-\text{tr}\left(\mathbf{n}^H\mathbf{s}\right)-\text{tr}\left(\mathbf{s}^H\mathbf{n}\right)+\text{tr}\left(\mathbf{s}^H\text{diag}\left(\mathbf{a}\right)\mathbf{P}^H\mathbf{H}^H\mathbf{n}\right)\nonumber\\
&+\text{tr}\left(\mathbf{n}^H\mathbf{H}\mathbf{P}\text{diag}\left(\mathbf{a}\right)\mathbf{s}\right)+\text{tr}\left(\mathbf{n}^H\mathbf{n}\right).\label{Squared Error Trace}
\end{align}
Taking the expected value of \eqref{Squared Error Trace} leads us to
\begin{align}
\mathbb{E}\left[\varepsilon\right]=&\text{tr}\left(\text{diag}\left(\mathbf{a}\right)\mathbf{P}^H\mathbf{H}^H\mathbf{H}\mathbf{P}\text{diag}\left(\mathbf{a}\right)\mathbf{R}_\mathbf{s}\right)\nonumber\\
&-\text{tr}\left(\mathbf{H}\mathbf{P}\text{diag}\left(\mathbf{a}\right)\mathbf{R}_{\mathbf{s}}\right)-\text{tr}\left(\text{diag}\left(\mathbf{a}\right)\mathbf{P}^H\mathbf{H}^H\mathbf{R}_{\mathbf{s}}\right)\nonumber\\
&+\text{tr}\left(\mathbf{R}_\mathbf{s}\right)+\text{tr}\left(\mathbf{R}_{\mathbf{n}}\right),\nonumber\\
=&\text{tr}\left(\text{diag}\left(\mathbf{a}\right)\mathbf{P}^H\mathbf{H}^H\mathbf{H}\mathbf{P}\text{diag}\left(\mathbf{a}\right)\right)-\text{tr}\left(\mathbf{H}\mathbf{P}\text{diag}\left(\mathbf{a}\right)\right)\nonumber\\
&-\text{tr}\left(\text{diag}\left(\mathbf{a}\right)\mathbf{P}^H\mathbf{H}^H\right)+N_r\left(1+\sigma_n^2\right),\label{Squared Error Expected value}
\end{align}
where the elements of the input vector are assumed uncorrelated with zero mean and unit variance. By taking the derivative of \eqref{Squared Error Expected value} with respect to power loading matrix $\mathbf{A}$ and using the equality $\frac{\partial \text{tr}\left(\mathbf{C}\mathbf{D}\right)}{\partial \mathbf{C}}=\mathbf{D}\odot \mathbf{I}$, where $\mathbf{C}$ is a diagonal matrix, we obtain
\begin{align}
\frac{\partial \mathbb{E}\left[\varepsilon\right]}{\partial \mathbf{A}}=&2\left(\mathbf{P}^H\mathbf{H}^H\mathbf{H}\mathbf{P}\text{diag}\left(\mathbf{a}\right)\right)\odot\mathbf{I}\nonumber\\
&-\left(\mathbf{P}^H\mathbf{H}^H\right)\odot\mathbf{I}-\left(\mathbf{H}\mathbf{P}\right)\odot\mathbf{I},\nonumber\\
=&2\left(\mathbf{P}^H\mathbf{H}^H\mathbf{H}\mathbf{P}\text{diag}\left(\mathbf{a}\right)\right)\odot\mathbf{I}-2\Re\left\{\left(\mathbf{H}\mathbf{P}\right)\odot\mathbf{I}\right\}.\label{Error derivative}
\end{align}
Employing a stochastic gradient descent approach we obtain the following update equation:
\begin{align}
\mathbf{a}\left[i\right]=&\mathbf{a}\left[i-1\right]-\mu\frac{\partial \mathbb{E}\left[\varepsilon\right]}{\partial \mathbf{A}}\nonumber\\
=&\mathbf{a}\left[i-1\right]-\mu\left(\mathbf{P}^H\mathbf{H}^H\mathbf{H}\mathbf{P}\text{diag}\left(\mathbf{a}\left[i-1\right]\right)\right)\odot\mathbf{I}\nonumber\\
&-\mu\Re\left\{\left(\mathbf{H}\mathbf{P}\right)\odot\mathbf{I}\right\},
\end{align}
where $\mu$ is the step size that governs the learning rate of the adaptive algorithm. The precoders in the previous equation are assumed to have columns with unitary norm. Moreover, the vector $\mathbf{a}$  is normalized before running the adaptive algorithm in order to have unitary norm. Therefore, the transmit power constraint is $\text{tr}\left(\text{diag}\left(\mathbf{a}\odot \mathbf{a}\right)\right)=1$. After each iteration the coefficients are properly scaled employing a power scaling factor $\beta$ to satisfy the transmit power constraint. Fig. \ref{C5 T1} shows the curves of \eqref{obejctive function 1} for three different linear precoders where only two streams are being transmitted.
In all cases the function is convex. Algorithm \ref{M-APA algorithm} summarizes the proposed adaptive power allocation strategy.

\section{Robust Adaptive Power Allocation}\label{c5 section robust power allocation}

Let us now derive the RM-APA algorithm, which takes into account the statistical knowledge of the CSIT imperfections. First, consider the square of the error function given by
\begin{align}
\varepsilon=&\lVert\mathbf{s}_p-\hat{\mathbf{H}}\mathbf{P}\text{diag}\left(\mathbf{a}\right)\mathbf{s}_p-\tilde{\mathbf{H}}\mathbf{P}\text{diag}\left(\mathbf{a}\right)\mathbf{s}_p-\mathbf{n}\rVert^2.
\end{align}
By expanding the terms, we have
\begin{align}
\varepsilon=&\mathbf{s}^H\text{diag}\left(\mathbf{a}\right)\mathbf{P}^H\hat{\mathbf{H}}^H\hat{\mathbf{H}}\mathbf{P}\text{diag}\left(\mathbf{a}\right)\mathbf{s}-2\Re\left(\mathbf{s}^H\hat{\mathbf{H}}\mathbf{P}\text{diag}\left(\mathbf{a}\right)\mathbf{s}\right)\nonumber\\
&+\mathbf{n}^H\mathbf{n}-2\Re\left(\mathbf{s}^H\tilde{\mathbf{H}}\mathbf{P}\text{diag}\left(\mathbf{a}\right)\mathbf{s}\right)-2\Re\left(\mathbf{s}^H\mathbf{n}\right)\nonumber\\
&+\mathbf{s}^H\text{diag}\left(\mathbf{a}\right)\mathbf{P}^H\tilde{\mathbf{H}}^H\tilde{\mathbf{H}}\mathbf{P}\text{diag}\left(\mathbf{a}\right)\mathbf{s}+\mathbf{s}^H\mathbf{s}\nonumber\\
&+2\Re\left(\mathbf{s}^H\text{diag}\left(\mathbf{a}\right)\mathbf{P}^H\hat{\mathbf{H}}^H\tilde{\mathbf{H}}\mathbf{P}\text{diag}\left(\mathbf{a}\right)\mathbf{s}\right)\nonumber\\
&+2\Re\left(\mathbf{s}^H\text{diag}\left(\mathbf{a}\right)\hat{\mathbf{H}}^H\mathbf{P}^H\mathbf{n}\right)\nonumber\\
&+2\Re\left(\mathbf{s}^H\text{diag}\left(\mathbf{a}\right)\tilde{\mathbf{H}}^H\mathbf{P}^H\mathbf{n}\right).
\end{align}
Including the trace operator over the right-hand side and taking the expected value we obtain
\begin{align}
\mathbb{E}_{\mathbf{s}|\mathbf{H}}\left[\varepsilon\right|\mathbf{H}]=&\text{tr}\left(\mathbf{R}_\mathbf{s}\right)-2\text{tr}\left(\Re\left(\hat{\mathbf{H}}\mathbf{P}\text{diag}\left(\mathbf{a}\right)\mathbf{R}_{\mathbf{s}}\right)\right)\nonumber\\
&+2\text{tr}\left(\Re\left(\text{diag}\left(\mathbf{a}\right)\mathbf{P}^H\hat{\mathbf{H}}^H\hat{\mathbf{H}}\mathbf{P}\text{diag}\left(\mathbf{a}\right)\mathbf{R}_\mathbf{s}\right)\right)\nonumber\\
&+\text{tr}\left(\mathbf{R}_{\mathbf{n}}\right)-2\text{tr}\left(\Re\left(\tilde{\mathbf{H}}\mathbf{P}\text{diag}\left(\mathbf{a}\right)\mathbf{R}_{\mathbf{s}}\right)\right)\nonumber\\
&+\text{tr}\left(\text{diag}\left(\mathbf{a}\right)\mathbf{P}^H\tilde{\mathbf{H}}^H\tilde{\mathbf{H}}\mathbf{P}\text{diag}\left(\mathbf{a}\right)\mathbf{R}_\mathbf{s}\right)\nonumber\\
&+\text{tr}\left(\text{diag}\left(\mathbf{a}\right)\mathbf{P}^H\hat{\mathbf{H}}^H\tilde{\mathbf{H}}\mathbf{P}\text{diag}\left(\mathbf{a}\right)\mathbf{R}_\mathbf{s}\right).\nonumber\\
\end{align}

Note that the system has only access to $\hat{\mathbf{H}}$. Moreover, the entries of $\tilde{\mathbf{H}}$ have a variance equal to $\sigma_e^2$, zero mean and are independent from the elements in $\hat{\mathbf{H}}$. By taking the expected value with respect to $\tilde{\mathbf{H}}$ to average out the effects of the channel uncertainties, we arrive at
\begin{align}
\mathbb{E}_{\tilde{\mathbf{H}}}\left[\varepsilon|\hat{\mathbf{H}}\right]=&\text{tr}\left(\mathbf{R}_\mathbf{s}\right)-2\text{tr}\left(\Re\left(\hat{\mathbf{H}}\mathbf{P}\text{diag}\left(\mathbf{a}\right)\mathbf{R}_\mathbf{s}\right)\right)\nonumber\\
&+\text{tr}\left(\text{diag}\left(\mathbf{a}\right)\mathbf{P}^H\hat{\mathbf{H}}^H\hat{\mathbf{H}}\mathbf{P}\text{diag}\left(\mathbf{a}\right)\mathbf{R}_\mathbf{s}\right)\nonumber\\
&+\text{tr}\left(\mathbf{R}_\mathbf{n}\right)+\text{tr}\left(\text{diag}\left(\mathbf{a}\right)\mathbf{P}^H\boldsymbol{\Xi}\mathbf{P}\text{diag}\left(\mathbf{a}\right)\mathbf{R}_\mathbf{s}\right),\label{Valor Esperado Error dada estimativa simplificado}
\end{align}
where we consider that $\tilde{\mathbf{h}}_k$ is a random vector independent from $\tilde{\mathbf{h}}_j$ with $j\neq k$. Furthermore, the diagonal error matrix $\boldsymbol{\Xi}$ is defined by
\begin{equation}
\boldsymbol{\Xi}=\mathbb{E}\left[\tilde{\mathbf{H}}^H\tilde{\mathbf{H}}\right]=\begin{bmatrix}
&\sigma_{e_1}^2 &0 &\cdots &0\\
&0 &\sigma_{e_2}^2 &\cdots &0\\
&\vdots &\vdots &\ddots &\vdots\\
&0 &0 &\cdots &\sigma_{e_{N_t}}^2
\end{bmatrix}.
\end{equation}
Without loss of generality, we consider that $\sigma_{e_i}^2=\sigma_{e_j}^2$ $\forall i,j$. Furthermore, $\sigma_{e_i}=N_r\sigma_e$, which leads us to 
\begin{equation}
\boldsymbol{\Xi}=N_r\begin{bmatrix}
&\sigma_{e}^2 &0 &\cdots &0\\
&0 &\sigma_{e}^2 &\cdots &0\\
&\vdots &\vdots &\ddots &\vdots\\
&0 &0 &\cdots &\sigma_{e}^2
\end{bmatrix}.
\end{equation}
By taking the derivative of \eqref{Valor Esperado Error dada estimativa simplificado} with respect to $\mathbf{A}$, we obtain
\begin{align}
\frac{\partial \mathbb{E}_{\tilde{\mathbf{H}}}\left[\varepsilon|\hat{\mathbf{H}}\right]}{\mathbf{A}}=&2\left(\mathbf{P}^H\hat{\mathbf{H}}^H\hat{\mathbf{H}}\mathbf{P}\text{diag}\left(\mathbf{a}\right)\right)\odot\mathbf{I}-2\Re\left(\left(\check{\mathbf{H}}\mathbf{P}\right)\odot\mathbf{I}\right)\nonumber\\
&+2\left(\mathbf{P}^H\boldsymbol{\Xi}\mathbf{P}\text{diag}\left(\mathbf{a}\right)\right)\odot\mathbf{I}\label{partial derivative A robust}
\end{align}
From \eqref{partial derivative A robust} we devise a gradient descent recursion, which is given by
\begin{align}
\mathbf{a}\left[i\right]=&\mathbf{a}\left[i-1\right]+\mu\Re\left(\left(\check{\mathbf{H}}\mathbf{P}\right)\odot\mathbf{I}\right)\nonumber\\
&-\mu\left(\mathbf{P}^H\hat{\mathbf{H}}^H\hat{\mathbf{H}}\mathbf{P}\text{diag}\left(\mathbf{a}\left[i-1\right]\right)\right)\odot\mathbf{I}\nonumber\\
&-\mu N_r\left(\left(\mathbf{P}^H\boldsymbol{\Xi}\mathbf{P}\text{diag}\left(\mathbf{a}\left[i-1\right]\right)\right)\odot\mathbf{I}\right)
\end{align}
The statistical information of the CSIT imperfection is included into the recursion of the power allocation coefficients, increasing the robustness against CSIT uncertainties. The proposed technique aims at maximizing the average sum-rate given a channel estimate $\hat{\mathbf{H}}$, since the the instantaneous rate is not achievable. 
\\
\begin{algorithm}[t!]
\SetAlgoLined
 given $\mathbf{H}$,$\mathbf{P}$ and $\mu$\;
 $\mathbf{a}\left[1\right]=\mathbf{0}$\;
 \For{$i=2$ \KwTo $I_t$}{
  $\frac{\mathbb{E}\left[\varepsilon\right]}{\partial \mathbf{A}}=2\left(\mathbf{P}^H\mathbf{H}^H\mathbf{H}\mathbf{P}\text{diag}\left(\mathbf{a}\left[i-1\right]\right)\right)\odot\mathbf{I}$\ $~~~~~~~~~-2\Re\left\{\left(\mathbf{H}\mathbf{P}\right)\odot\mathbf{I}\right\}$\;
  
  $\mathbf{a}\left[i\right]=\mathbf{a}\left[i-1\right]-\mu\frac{\partial \mathbb{E}\left[\varepsilon\right]}{\partial \mathbf{A}}$\;
  \If{$\textrm{\rm tr}\left(\textrm{\rm diag}\left(\mathbf{a}\left[i\right]\odot \mathbf{a}\left[i\right]\right)\right)\neq 1$}{
   $\beta=\sqrt{\frac{1}{\textrm{tr}\left(\textrm{diag}\left(\mathbf{a}\left[i\right]\odot \mathbf{a}\left[i\right]\right)\right)}}$\;
   $\mathbf{a}\left[i\right]=\beta\mathbf{a}\left[i\right]$\;
   }
 }
 \caption{MMSE Adaptive Power allocation}
 \label{M-APA algorithm}
\end{algorithm}\vspace{1em}

\begin{algorithm}[t!]
\SetAlgoLined
 given $\hat{\mathbf{H}}$, $\mathbf{P}$, $\boldsymbol{\Xi}$ and $\mu$\;
 $\mathbf{a}\left[1\right]=\mathbf{0}$\;
 \For{$i=2$ \KwTo $I_t$}{
  $\frac{\partial \mathbb{E}_{\tilde{\mathbf{H}}}\left[\varepsilon|\hat{\mathbf{H}}\right]}{\mathbf{A}}=2\left(\mathbf{P}^H\hat{\mathbf{H}}^H\hat{\mathbf{H}}\mathbf{P}\text{diag}\left(\mathbf{a}\right)\right)\odot\mathbf{I}-2\Re\left(\left(\hat{\mathbf{H}}\mathbf{P}\right)\odot\mathbf{I}\right)$\
  $~~~~~~~~~~~~~~+2\left(\mathbf{P}^H\boldsymbol{\Xi}\mathbf{P}\text{diag}\left(\mathbf{a}\right)\right)\odot\mathbf{I}$\;
  
  $\mathbf{a}\left[i\right]=\mathbf{a}\left[i-1\right]-\mu\frac{\partial \mathbb{E}_{\tilde{\mathbf{H}}}\left[\varepsilon|\hat{\mathbf{H}}\right]}{\partial \mathbf{A}}$\;
  \If{$\textrm{\rm tr}\left(\textrm{\rm diag}\left(\mathbf{a}\left[i\right]\odot \mathbf{a}\left[i\right]\right)\right)\neq 1$}{
   $\beta=\sqrt{\frac{1}{\textrm{tr}\left(\textrm{diag}\left(\mathbf{a}\left[i\right]\odot \mathbf{a}\left[i\right]\right)\right)}}$\;
   $\mathbf{a}\left[i\right]=\beta\mathbf{a}\left[i\right]$\;
   }
 }
 \caption{MMSE Robust Adaptive Power allocation}
\end{algorithm}\vspace{1em}

\section{Simulations}\label{section Simulations}

In this section, the performance of the proposed power allocation techniques is assessed against conventional approaches. We consider a MU-MIMO system where the BS is equipped with four antennas and transmits data to two users, each equipped with two antennas. The inputs are statistically independent and follow a Gaussian distribution. A flat fading Rayleigh channel, which remains fixed during the transmission of a packet, is considered. Moreover, we assume additive white Gaussian noise with zero mean and unit variance. It follows that the SNR varies with $E_{tr}$.   

\begin{figure}[h]
\begin{center}
\includegraphics[scale=0.4]{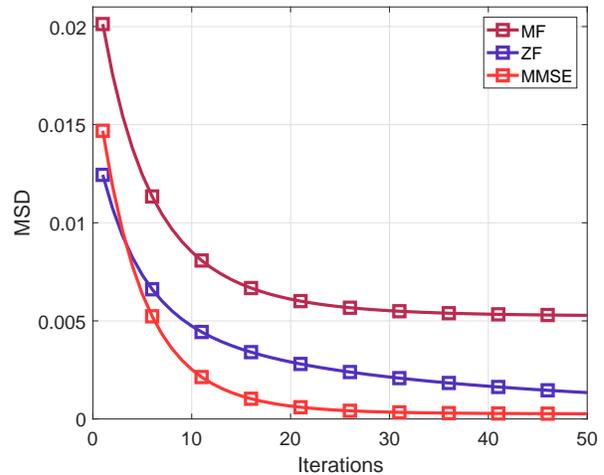}
\vspace{-1.5em}
\caption{Learning curves of the adaptive power allocation techniques.}
\label{C5 Figure00}
\end{center}
\end{figure}

First, let us analyse the learning curves of the adaptive algorithms. Fig \ref{C5 Figure00} shows the mean square deviation (MSD) obtained with three different linear precoders, namely the matched filter (MF), the zero-forcing (ZF) and the MMSE precoders \cite{Joham2005}. To compute the MSD, we employ the optimum value that solves \eqref{obejctive function 1}. This value was obtained through exhaustive search with a step of $0.005$. The learning curves were obtained by averaging over 1000 independent Monte Carlo simulations. The step of the adaptive algorithm was set to $0.01$ for all precoders. The adaptive algorithm reaches its steady state with about $30$ iterations, which corresponds to a fast convergence.

\begin{figure}[h]
\begin{center}
\includegraphics[scale=0.4]{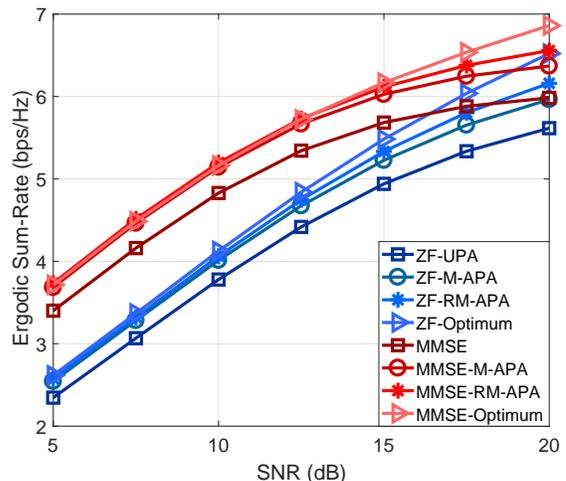}
\vspace{-1.5em}
\caption{Sum-rate performance with linear precoding scheme, $N_t=4$, $N_k=2$, $K=2$, and $\sigma_e^2=0.1$.}
\label{C5 Figure1}
\end{center}
\end{figure}

In the next experiment, we consider an imperfect CSIT scenario with $\sigma_e^2=0.1$. The ergodic sum-rate was obtained by averaging 10000 independent channels. Fig. \ref{C5 Figure1} shows the performance obtained employing different power allocation techniques with ZF and MMSE precoders. As expected, the best performance is attained with the exhaustive search, i.e., ES. However, the very high computational complexity of the ES approach makes it impractical. Moreover, the time spent increases exponentially with the number of users. The proposed strategies not only increase the performance of the system when compared to UPA but also have low computational complexity, which is very important for real communication systems. We can notice that the robust RM-APA approach performs better than the M-APA algorithm at the expense of a slight increase in computational complexity, which is justified based on the improved performance of RM-APA over M-APA. In addition, the proposed M-APA and RM-APA algorithms significantly outperform the uniform power allocation, i.e., UPA, and the random power allocation, denoted as Random, strategies.

\section{Conclusion}\label{section conclusion}
In this paper, the M-APA and RM-APA adaptive power allocation algorithms were developed and shown to obtain  better performance than the conventional UPA under imperfect CSIT. Recursive expressions to update the power allocation parameters were derived, which keep linear complexity since only simple multiplications and additions are required. RM-APA employs statistical information from the error, attaining the best performance among the proposed algorithms.






%



\bibliographystyle{IEEEbib}
\bibliography{Powerallocation}

\end{document}